\documentclass[reprint, amsmath,amssymb,aps,pra]{revtex4-1}
\usepackage{feynmp}
\usepackage{subcaption}

\usepackage{graphicx}% Include figure files
\usepackage{dcolumn}% Align table columns on decimal point
\usepackage{bm}% bold math
\begin{document}

%\preprint{APS/123-QED}

\title{One Leptoquark to unify them? Neutrino masses and unification in the light of $(g-2)_\mu$, $R_{D^{(\star)}}$ and $R_K$ anomalies}% Force line breaks with \\

\author{Oleg Popov}%
 \email{opopo001@ucr.edu}
\author{G A White}%
 \email{graham.white@monash.edu}
\affiliation{
${}^*$Department of Physics and Astronomy, \\ University of California, Riverside, California 92521, USA \\
${}^\dagger$School of Physics and Astronomy, Monash University, Victoria 3800, Australia
\\
ARC Center of Excellence for Particle Physics at the Tera-scale, Monash University, Victoria 3800 Australia}

\begin{abstract}
Leptoquarks have been proposed as a possible explanation of anomalies in $\bar{B}\mapsto D^{*} \tau \bar{\nu } $ decays, the apparent anomalies in $(g-2)_\mu$ experiments and a violation of lepton universality. Motivated by this, we examine other motivations of leptoquarks: radiatively induced neutrino masses in the presence of a discrete symmetry that prevents a tree level see-saw mechanism, gauge coupling unification, and  vacuum stability at least up to the unification scale. We present a new model for radiatively generating a neutrino mass which can significantly improve gauge coupling unification at one loop. We discuss this, and other models in the light of recent work on flavour anomalies. 
%\begin{description}
%\verb+\pacs{#12.10.Kt,#14.60.Pq,#14.80.Sv}
%\end{description}
\end{abstract}

\pacs{Valid PACS appear here}% PACS, the Physics and Astronomy
\maketitle

\section{introduction}

Recently there has been a lot of interest in leptoquarks as a possible explanation for some striking deviations from the standard model \cite{explanation1,explanation2,explanation3,explanation4,explanation5,explanation6,explanation7,explanation8,Allanach:2015ria,Queiroz:2014pra,Calibbi:2015kma,Deppisch:2016qqd} including anomalous B decays observed in BaBar \cite{anomalousBdecays1,anomalousBdecays2}, Belle \cite{anomalousBdecays3} and LHCb \cite{anomalousBdecays4,anomalousBdecays5,anomalousBdecays6}, a violation in lepton universality \cite{leptonuniversality} and a deviation from the standard model prediction of $(g-2)_\mu$ \cite{muonmag}. A particularly interesting claim was that all three anomalies could be explained via the addition of a single leptoquark \cite{explanation4}.  Specifically they chose a single $\sim $ TeV scale leptoquark, $\phi$, that is a colour triplet and an SU(2)$_L$ singlet with hyper charge $-1/3$. The correction to the standard model Lagrangian due to the existence of this leptoquark is
Lagrangian
    \begin{align}
        {\cal L} & \ni (D_\mu \phi) ^\dagger  D ^\mu \phi - M_{\phi }^2 \phi ^2 -g_{h \phi }|\Phi |^2|\phi |^2 \nonumber \nonumber \\
        &+ \bar{Q} ^c \lambda _L i \tau _2 L \phi ^\star + \bar{u} ^c _R \lambda _R e_R \phi ^\star + {\rm h.~c.~} \label{lagrangian}
    \end{align}
    where $\Phi $ is the standard model Higgs doublet. We will denote such a leptoquark by its gauge quantum numbers $(3,1,-1/3)$. The attempt to explain lepton universality with such a leptoquark was brought into significant doubt recently \cite{Becirevic:2016oho} (who explain the anomaly with a leptoquark with quantum numbers $(3,2,1/6)$). However it has been recently demonstrated that the  $(3,1,-1/3)$ leptoquark indeed can explain all three anomalies \cite{Cai:2017wry}. 
\\
Leptoquarks are also of theoretical interest as the tend to appear in various grand unified theories (GUTs). Indeed the particular choice of leptoquark used in ref. \cite{explanation4} appears in E$6$ GUTs \cite{E6ssm1,E6ssm2} and the $(3,2,1/6)$ leptoquark arises for example in an SU(5) GUT. Leptoquarks have also been proposed as a non-supersymmetric catalyst of gauge coupling unification \cite{unification1,unification2,unification3,unification4} and a cause of neutrino masses via radiative corrections \cite{unification3,LQmass2,LQmass3,LQmass4,LQmass5,LQmass6,LQmass7,LQmass8,LQmass9,LQmass10,LQmass11,LQmass12}. These explanations of the neutrino mass are phenomenologically attractive as unlike the see saw mechanism, these models have predictions at much lower energies. 
\\
It is of interest to us whether using these leptoquarks to explain these flavour anomalies is compatible with some of these theoretical motivations and if not what is the minimal extension needed for such compatibility. Specifically we look at radiatively induced neutrino masses, gauge coupling unification, and vacuum stability. We will consider both the $(3,1,-1/3)$ and $(3,2,1/6)$ representations and propose a model for each that radiatively induces Weinberg dim-5 operator ($L L H H /\Lambda$) to produce Majorana neutrino mass at one and two loops respectively. This requires the introduction of new particles for both cases. For the 1-loop case the addition of heavy d-type quarks is needed, 1 per family. We show a sample parameter space that avoids experimental bounds and gives neutrino masses of the correct order. As for 2 loop scenario with $(3,1,-1/3)$ leptoquark we include heavy scalar diquarks with different isospin structure for the loop completion. Proton decay is also of interest when one considers unification. However, the $(3,2,1/6)$ easily avoids constraints on proton decay and the $(3,1,-1/3)$ Leptoquarks are similarly safe in the presence of a discrete symmetry where SM leptons and the leptoquark have opposite parity.

For gauge coupling unification we find two paths - first through the $(3,2,1/6)$ case without the extra particles required to radiatively induce a neutrino mass - and second through the introduction of both the $(3,1,-1/3)$ leptoquark and new particles required to generate the neutrino mass at one loop. A third more difficult path also becomes manifest in the case of a $(3,1,-1/3)$ leptoquark through large leptoquark couplings. However, this puts constraints on the leptoquark couplings which partly conflict with the constraints on leptoquark coupling given in \cite{explanation4} as they had $\lambda ^{R}_x<< \lambda ^L _x$ which is generically the opposite condition to what gauge coupling unification requires.  One in principle has enough freedom left to achieve gauge coupling unification although this requires some couplings to be brought close to the perturbativity bound. This makes the calculation susceptible to high theoretical error.  Most of this tension is due to one of the anomalies - the observed violation of lepton universality. This is precisely the anomaly that is difficult to explain via the $(3,1,-1/3)$ leptoquark \cite{Becirevic:2016oho}.
By contrast, vacuum stability is substantially boosted due to the improved running of the Higgs quartic coupling when the portal coupling is non-negligible. This is true for both leptoquark representations. Furthermore the extra fermions used to generate a neutrino mass at $1$ loop will only contribute to the running of the Higgs quartic at two loop. For the case where gauge coupling unification is achieved through large leptqoaruk couplings one requires a larger Higgs portal coupling to maintain stability. 
\\
The structure of this paper is as follows. We discuss minimal modifications to achieve neutrino masses aside from the tree level term in section \ref{sec:neutrino}. In sections \ref{sec:GUT} and \ref{sec:vacuum} we discuss the modification to the running of the gauge coupling constants and the Higgs quartic coupling respectively, thereby discussing the viability of gauge coupling unification and vacuum stability in this model. We then discuss how compatible these constraints are with collider constraints and the need to simultaneously explain violations of lepton universality, anomalies in $B$ decays and the measurement of $(g-2)_\mu$ experiments in section \ref{sec:pheno}.  We conclude in section \ref{sec:conclusion} 

\section{Possible Extensions of the Leptoquark model for the Neutrino Mass Generation}\label{sec:neutrino}
To generate Neutrino mass using the leptoquark model discussed above, we need to go beyond this model and include more particles/fields. Depending on the content added, a neutrino mass can be generated at 1 or 2 loop level. There are several examples in the literature that generate neutrino mass through the use of multiple leptoquarks fields \cite{unification3,LQmass2,LQmass3,LQmass4,LQmass5,LQmass6,LQmass7,LQmass8,LQmass9,LQmass10,LQmass11,LQmass12}. If Dirac Neutrino Mass is generated then potentially additional symmetries, global or gauged, are needed to forbid tree level Neutrino mass generation. Here we briefly discuss 2 extensions of the leptoquark model to generate the effective Majorana neutrino mass at the 1 and 2 loop order through an effective d-5 Weinebrg operator\cite{weinberg} $ L L H H/\Lambda$ with only one leptoquark field in the model.  
\subsection{2 loop Majorana Neutrino Mass}
\begin{figure}[h]
\centering
\begin{subfigure}[h]{0.5\textwidth}
\includegraphics[scale=1,trim={1cm 23cm 11cm 1cm},clip]{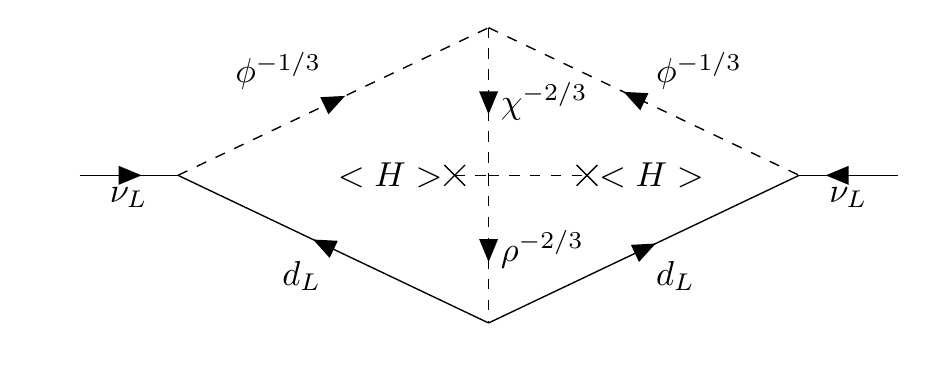}
%\caption{2 loop Majorana Neutrino Mass}
\label{fig:Mnu2l1}
\end{subfigure} \\
\begin{subfigure}[h]{0.5\textwidth}
\includegraphics[scale=1,trim={1cm 23cm 11cm 1cm},clip]{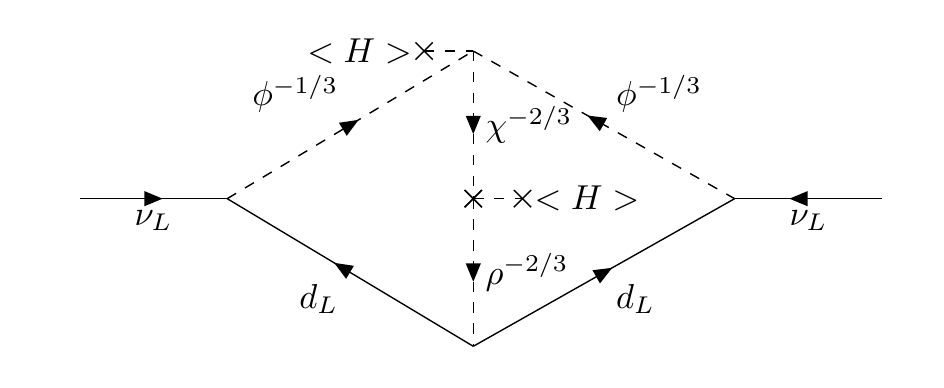}
%\caption{2 loop Majorana Neutrino Mass}
\label{fig:Mnu2l2}
\end{subfigure}
\caption{2 loop Majorana Neutrino Mass through effective d-5 Weinberg operator.}
\label{fig:2loopMnu}
\end{figure}
Generating a Majorana neutrino mass at 2 loop order from the leptoquark model mentioned in the present work requires the addition of extra colored charged scalar multiplets. Figure \ref{fig:2loopMnu} shows 2 different variants of Majorana neutrino mass with an effective Weinberg operator. In the first diagram, external Higgs fields are coupled symmetrically at the center vertex to the charged scalar line, and another one having one of Higgs field coupled to the leptoquarks at the top vertex and the second Higgs field coupled at the center. Both diagrams require the LQ$\phi$ coupling, fields added to complete the loops do not couple to the Leptons and Quarks simultaneously and so do not effect the results of the anomaly calculations done in \cite{explanation4} at the leading order calculations. The first diagram requires the minimal addition of 2 diquarks, which have the following representations (6,1, -2/3), (6,3,-1/3) which are all color sextets. The charged scalar field coupled to the leptoquarks must be an iso-singlet where as the charged scalar field coupled to SM quark doublets is a triplet under SU(2)$_L$. Lepton number is broken at the soft dim-3 $\phi \phi \chi^*$ term by 2 units. The second diagram of Fig.\ref{fig:2loopMnu} is similar to the first diagram but now $\chi\sim$(6,2,1/6) is a doublet under SU(2)$_L$ and $\rho\sim$(6,3,2/3) is an-iso triplet as in the first diagram. The Lepton number is broken at the only dim-3 soft term by 2 units.

If the $\phi\sim$(3,2,1/6) leptoquark representation is used instead of (3,1,-1/3), then we get $\bar{d}_R L \phi$ operator instead of $LQ\phi$. Then the required fields for the first 2-loop completion are $\chi\sim$(6,3,1/3) and $\rho\sim$(6,1,2/3). For the second 2-loop diagram the required scalar diquark fields are $\chi\sim$(6,2,7/6) and $\rho\sim$(6,1,2/3).

%\hrule
\subsection{1 loop Majorana Neutrino Mass through Lepotoquark and $d_R$ mixing}
\begin{figure}[h]
\centering
\includegraphics[scale=1,trim={1cm 22cm 13cm 1cm},clip]{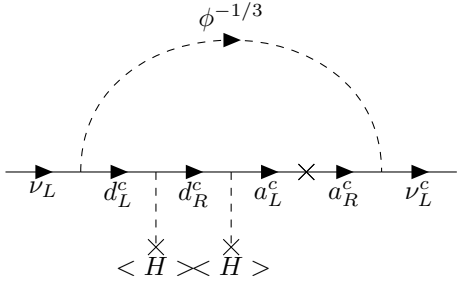}
\caption{1 loop Neutrino Mass Diagram through d quark mixing with effective d-5 Weinberg operator.}
\label{fig:daMnu}
\end{figure}
Here we briefly discuss another extension of the leptoquark model mentioned above to generate Majorana Neutrino Mass at 1 loop level shown in Fig. \ref{fig:daMnu}. Table \ref{table:tab1} shows the field contents of the model. Besides the SM and a leptoquark field, we add 3 generations of heavy vectorlike quark doublets with a hypercharge -5/6 to to complete the loop. The new Lagrangian terms are shown below. To break the Lepton number softly we need the right-handed heavy quark fields to carry 2 units of L number and the Left-handed heavy quark fields to carry no L number. 
\begin{table}
\begin{center}
%\centering
\begin{tabular}{|c|c|c|c|c|}
\hline
Particle & SU(3)$_c$ & SU(2)$_L$ & U(1)$_Y$ & Flavour \\ \hline
Q & 3 & 2 & 1/6 & 3 \\
d$_R^c$ & 3$^*$ & 1 & +1/3 & 3 \\
u$_R^c$ & 3$^*$ & 1 & -2/3 & 3 \\
L & 1 & 2 & -1/2 & 3 \\
e$_R^c$ & 1 & 1 & +1 & 3 \\
A$_{R,L}$ & 3 & 2 & -5/6 & 3 \\
H & 1 & 2 & 1/2 & 1 \\
$\phi$ & 3 & 1 & -1/3 & 1 \\ \hline
\end{tabular}
\caption{\raggedright{}Particle content of the model generating Neutrino mass at 1 loop order.}
\label{table:tab1}
\end{center}
\end{table}
\begin{align}
\mathcal{L}_{new,4D}^Y\,&\subset \, y_1\, \overline{Q_L^c} L \phi^* + y_2\, \overline{u_R^c} e_R \phi^* \nonumber \\ & + y_3\, \overline{A_R} L \phi + y_{\epsilon} \overline{d_R} A_L H + h.c. \\
\mathcal{L}_{3D}\,&\subset\,\mathcal{M}_A \overline{A} A \\
V(H,\phi)\,&=\, -m_1^2 \left|H\right|^2 + \frac{\lambda_1}{4} \left|H\right|^4 + m_2^2 \left|\phi\right|^2 \nonumber \\ & + \frac{\lambda_2}{4} \left|\phi\right|^4 + g_{h \phi} \left(H^{\dagger}H\right)\left|\phi\right|^2
\end{align}
%\mathcal{L}_{\Delta L \neq 0}\,=\, \overline{Q_L^c} Q_L \phi + \overline{d_R} u_R^c \phi^* + \text{h.c.} 
The y$_{\epsilon}$ Yukawa term is the mixing of SM d type quarks and heavy quarks and required to be small. The effective mass matrix for the d type quark mixing is shown below, where the top left entry is the usual Higgs mass of the SM and the bottom right entry is the invariant mass of the heavy quarks, where as the off-diagonal term is the source of mixing of SM quarks and the heavy quarks. 
\begin{align}
\mathcal{L_{\text{eff mix}}}\,&=\,
\overline{\left(\begin{matrix}
d_L & a_L^{-1/3}
\end{matrix}\right)} \mathcal{M_{\text{da}}} 
\left(\begin{matrix}
d_R \\
a_R^{-1/3}
\end{matrix}\right) + h.c. \\
\mathcal{M_{\text{da}}}\, &=\,
\left(\begin{matrix}
y_d \nu & 0 \\
y_{\epsilon} \nu & M_A \\
\end{matrix}\right)
\end{align}

Diagonalizing this mass matrix, we obtain the following mass eigenstates

\begin{align}
    m_{D1} &= c_L c_R y_d \nu/\sqrt{2} + s_L c_R y_{\epsilon} \nu/\sqrt{2} + s_L s_R M_A, \\
    m_{D2} &= c_L c_R M_A + s_L s_R y_d \nu/\sqrt{2} - c_L s_R y_{\epsilon} \nu/\sqrt{2},
\end{align}

where the c/s stand for Cos/Sin and L/R subscripts stand for Left/Right mixing angles of the Left/Right chiral components of the fermion fields. The mixing is given by following relations

\begin{align}
    \left(\begin{matrix}D_{1L} \\ D{2L}\end{matrix}\right)&=U_L\times \left(\begin{matrix}d_{L} \\ a_L^{-1/3}\end{matrix}\right)\\
    \left(\begin{matrix}D_{1R} \\ D{2R}\end{matrix}\right)&=U_R\times \left(\begin{matrix}d_{R} \\ a_R^{-1/3}\end{matrix}\right)\\
    tan(2 \theta_L) &= \frac{2 \gamma^2 y_d y_{\epsilon}}{\gamma^2 \left(y_d^2-y_{\epsilon}^2\right)-1}\\
    tan(2 \theta_R) &= \frac{2 \gamma y_{\epsilon}}{\gamma^2 \left(y_d^2+y_{\epsilon}^2\right)-1}\\
    &\text{with }\gamma = \frac{\nu}{\sqrt{2}M_A}.
\end{align}

\begin{figure}[h]
\centering
\includegraphics[scale=0.7,trim={0cm 0cm 0cm 0cm},clip]{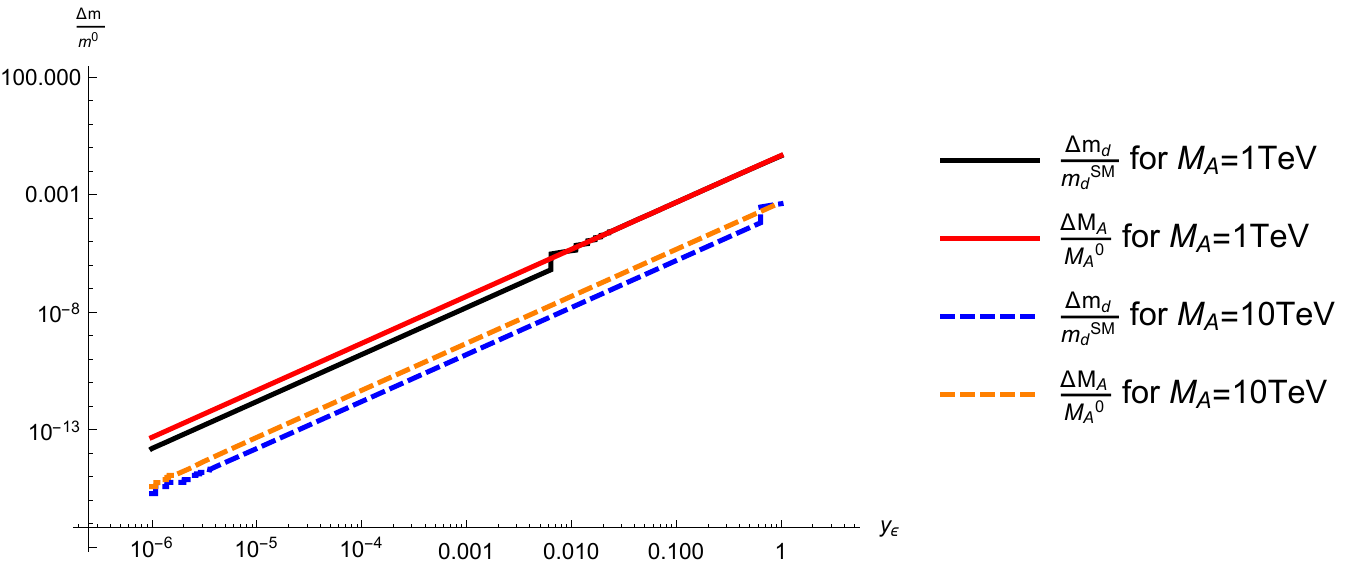}
\caption{Deviation of 2 mass eigenvalues from their unmixed values for $M_A$=1TeV and $M_A$=10TeV with Yukawa's set to 1 and m$_{\phi}$=1TeV.}
\label{fig:MD12Fig}
\end{figure}

Fig.~\ref{fig:MD12Fig} shows the deviations of mass eigenstates, light d quark and heavy a$^{-1/3}$ state, from their unmixed values for different values of y$_{\epsilon}$ and $M_A$ for Yakawa's set to 1 and $m_{\phi}$=1TeV.

1 loop radiative neutrino mass given in Fig. \ref{fig:daMnu} can be evaluated to 

\begin{align}
    m_{\nu}&=\frac{y_1 y_3 m_{\phi} s_R c_L}{16 \pi^2} x_1 \left[ 2 log\left(\frac{x_2}{x_1}\right) + f(x_2) - f(x_1)\right]\\
    &\text{where } x_i=m_{Di}/m_{\phi} \text{ and } f(x)=\frac{log(x)}{x^2-1}
\end{align}
\begin{center}
\begin{figure}[h]
%\centering
\includegraphics[scale=0.8,trim={0cm 0cm 0cm 0cm},clip]{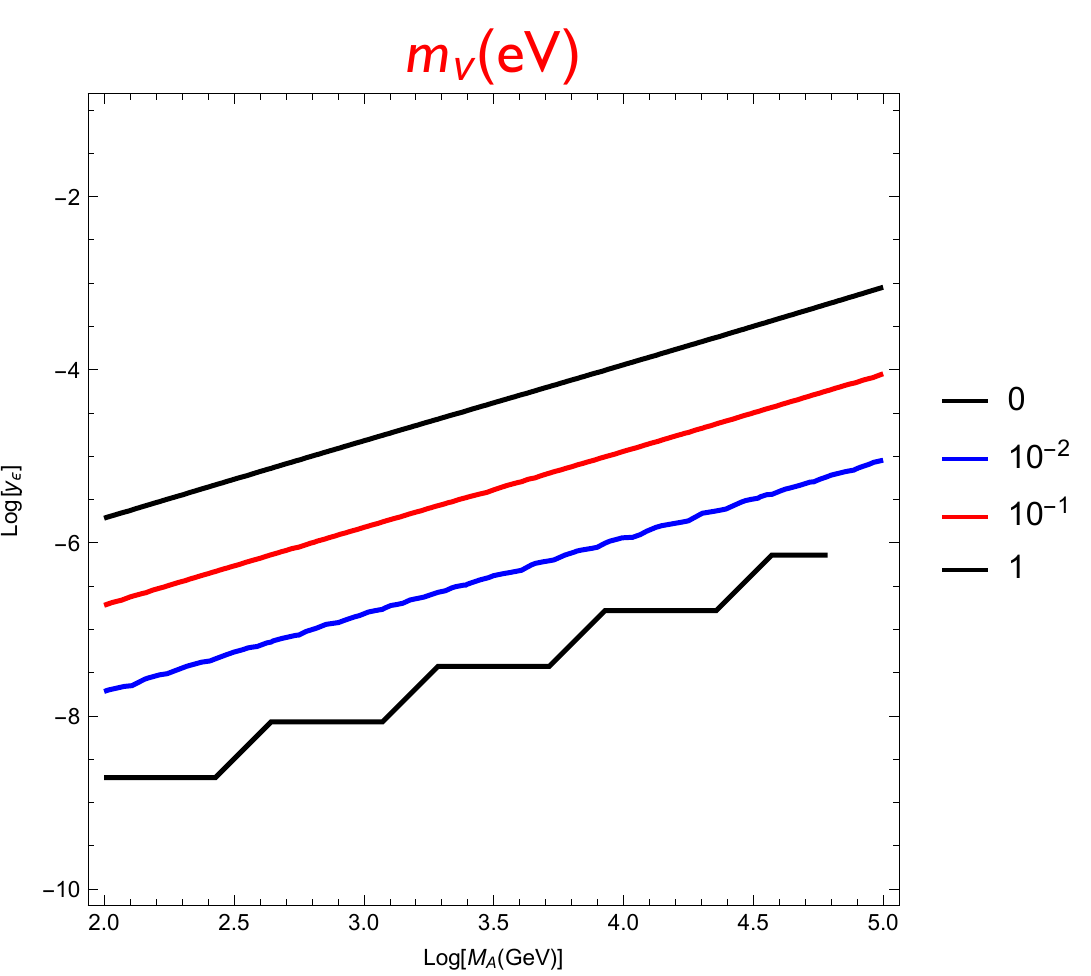}
\caption{Neutrino mass spectrum for different values of y$_{\epsilon}$ and M$_A$ with Yukawa's set to 1 and m$_{\phi}$=1TeV.}
\label{fig:mnuFig}
\end{figure}
\end{center}
Fig.~\ref{fig:mnuFig} shows a sample parameter space that gives the correct neutrino mass order for different values of $y_{\epsilon}$ and $M_A$, Yukawa's are set to 1, $m_{\phi}$=1TeV. 

The heavy quark searches set limits on the mixing of d-type quarks and the masses of the heavy quarks. Leptoquark searches set limits on the leptoquark couplings to LQ and Leptoquark masses. Possible CP violating phases might occur in the 6$\times$6 d-type mass matrix.

\section{Gauge Coupling Unification}\label{sec:GUT}
When one examines how the three gauge coupling constants of the standard model run, they come close to intersecting at a high energy scale. It is of course now known that they do not unify without some new physics entering in at a lower scale, often at a testable scale just above the weak scale.  The Leptoquarks change the running of gauge coupling constants and it was first suggested that a weak scale Leptoquark could cause gauge coupling unification to manifest in ref. \cite{unification1} and as mentioned in the introduction this has become a major motivation and attraction for considering low scale leptoquarks. It is therefore of interest whether this is feasible in a model where a leptoquark is used to simultaneously explain flavour anomalies and neutrino masses. 

We will find that by itself the $(3,1,-1/3)$ representation makes gauge coupling unification worse unless the leptoquark couplings are large enough to make a difference at two loops. This however puts criteria on the leptoquark couplings that are in tension with both perturbativity and constraints needed to explain flavour anomalies. This is also in contrast with the $(3,2,1/6)$ leptoquark. It was recently shown that lepton flavour universality and gauge coupling unification are compatible via the $(3,2,16)$ leptoquark \cite{Cox:2016epl}. However, when one adds the particles required to radiatively generate a neutrino mass, we find that such particles tend to make unification worse for the $(3,2,1/6)$ leptoquark. In contrast the additional fermions required to generate a neutrino mass at one loop, $A_{L,R}$, can easily improve gauge coupling unification for the $(3,1,-1/3)$ representation.

The one loop beta functions changes when one extends the standard model by a leptoquark. Defining the parameter $b_i$ as
\begin{equation}
    \frac{\partial g_i}{\partial \mu} =- \frac{b_i}{16 \pi ^2} g_i ^3
\end{equation}
it is straightforward to derive a measure of unification \cite{Aitchison:2005cf}
\begin{equation}
    \delta U = \frac{\alpha _3 ^{-1} (\mu )-\alpha _2 ^{-1} (\mu )}{\alpha _2 ^{-1} (\mu )-\alpha _1 ^{-1} (\mu )} - \frac{b_2 - b_3}{b_1-b_2}
\end{equation}
where a low value of $|\delta U|$ indicates one is close to unification. Choosing $\mu = 1$ TeV one has for the standard model $\delta U=0.19$. It is of course well known that the $(3,2,1/6)$ leptoquark improves unification. Introducing such leptoquarks at $1$ TeV and running the standard model couplings up to the TeV scale using the numerical package SARAH \cite{sarah} gives a value of $\delta U= (0.13,0.06,-0.02)$ for $(1,2,3)$ generations respectively. On the other hand, adding the leptoquark defined in equation \ref{lagrangian} modifies these values of $b_i$ such that $b_1=-25/6$ $b_3=41/6$ and $b_2$ remains unchanged. Again running the standard model couplings up to the TeV scale one finds that for the leptoquark extension one has $\delta U=0.21$, worse than the standard model. The reason for this is that one needs to change $b_2$ faster than one changes $\frac{1}{2}(b_1+b_3)$ in order to achieve unification but the leptoquark is an isospin singlet. 

\begin{figure}

        \includegraphics[width=0.9\linewidth]{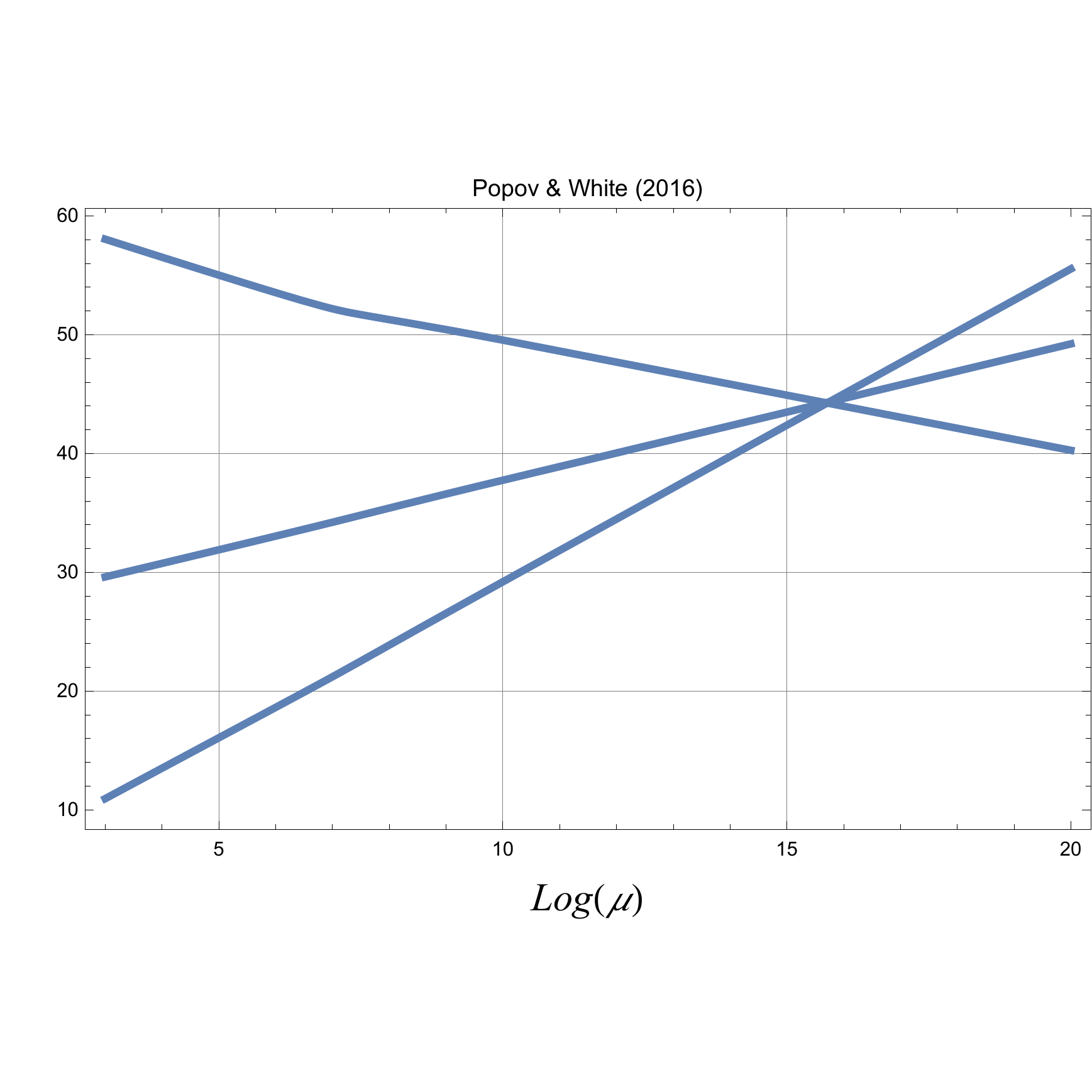}\caption{\raggedright{}Gauge coupling unification($\alpha^{-1}$) achieved via strong right handed leptoquark couplings. With a single generation the couplings get precariously close to the perturbativity bound, a problem accentuated if one wishes to explain a violation in lepton univerality, adding in a great deal of theoretical uncertainty which can be alleviated with extra generations of leptoquarks.}
    \label{Fig:Gauge}
    
\end{figure}

In principle the 2-loop corrections to the beta functions can drive gauge coupling unification. The leptoquark couplings $\lambda _{L}$ and $\lambda _R$ affect the two loop beta functions for gauge coupling constants at two loop level as follows

\begin{align}
    \beta_{g_1} & \ni \frac{1}{(4 \pi)^2} \frac{25 g_1^3}{6} - \frac{1}{(4 \pi)^4} \frac{(30  {\rm Tr}[\lambda _{L}\lambda _L ^\dagger] -78 {\rm Tr}[\lambda _{R}\lambda _R ^\dagger])}{30} g_1^3 \nonumber \\
    \beta_{g_2} & \ni - \frac{1}{(4 \pi)^2} \frac{19 g_2^3}{6}  - \frac{1}{(4 \pi)^4} (3 {\rm Tr}[\lambda _{L}\lambda _L ^\dagger] ) g_2^3 \nonumber  \\
    \beta_{g_3} & \ni - \frac{1}{(4 \pi)^2} \frac{41g_3^3}{6} - \frac{1}{(4 \pi)^4} \frac{(30 {\rm Tr}[\lambda _{L}\lambda _L ^\dagger] +15 {\rm Tr}[\lambda _{R}\lambda _R ^\dagger])}{30} g_3^3. 
\end{align}
Note that the Higgs portal coupling does not effect the running. The Higgs portal coupling is the major player in achieving a boost to the stability of the vacuum so two constraints - vacuum stability and gauge coupling unification - are only moderately correlated. Typically raising the left handed coupling constants takes one further away from coupling unification whereas raising the right handed coupling gets one closer. If all left handed couplings are null then gauge coupling unification is achieved when \begin{equation} {\rm Tr} \left[ \lambda _R \lambda _R ^\dagger \right] \sim 4. \end{equation} This number rises if the left handed leptoquark couplings rises. 
\\
If the leptoquark is used to explain the apparent violation of lepton universality, it turns out to be difficult to avoid at least one leptoquark coupling being large, namely the $\lambda ^L _{c \mu }$ coupling. This raises the required value of the right handed leptoquark coupings. Fig. \ref{Fig:Gauge} shows the running of the gauge couplings for leptoquark couplings chosen to satisfy all phenomenology constraints as well as explaining all three flavour anomalies. As before, we take all standard model couplings at weak scale and use the standard model RGEs defined to two loop to run all parameters from the weak scale to the mass of the leptoquark (which we set to be $1$ TeV). In principle, there is enough freedom left over from these constraints to achieve gauge coupling unification if some couplings are made sufficiently large. The results however should be taken with a heavy grain of salt because the running of the leptoquark couplings are such that they increase and approach the perturbativity bound. Therefore the calculation is expected to have a high amount of uncertainty.

Next let us turn to the extra particle content required to produce a neutrino mass at one and two loop respectively. For the two loop case we need two additional particles including an isospin triplet. This is true for both representations of leptoquark that we consider. This particle makes too dramatic a change to $b_2$ to assist unification unless its mass is quite high $\approx {\cal O} (10^{13})$ GeV. For the case where one induces a neutrino mass at one loop, the new particles change the values of $b_i$ as follows
\begin{align}
    \delta b_1 &= \left( -\frac{5}{6} \right) N_g \\
    \delta b_2 &= \left(-2 \right) N_g  \\
    \delta b_3 &= \left(- \frac{4}{3} \right) N_g
\end{align}
where $N_g$ is the number of generations active at that mass scale. For a single generation active from $1$ TeV up to around the GUT scale one essentially has unification $\delta U=0.015$. Similarly one achieves unification if two generations of $A_{L,R}$ with masses around $10-1000$ TeV and the third generation near the GUT scale. The masses of the lightest two generations can be within a wide range of values however. Varying their masses between $10$ and $1000$ TeV one finds that $|\delta U|$ ranges between $0.05 \lesssim |\delta U| \lesssim 0.08$. If one has three generations of these fermions it is unavoidable that at least one generation is very heavy. By contrast, the analogous fermions used to generate a neutrino mass at one loop for the $(3,2,1/6)$ leptoquark makes gauge coupling unification worse because they are isospin singlets.

Let us conclude this section with a summary of this section. The extra A particles make unification at 1 loop fairly straightforward for the (3,1,1/6) model. Without such particles one can in principle achieve unification at 2 loops with very large couplings, a result that is in some tension with flavour anomalies but not stability. By contrast, while the (3,2,1/6) model achieves unification without a radiatively induced neutrino mass, introducing such extra particles is in tension with unification.

\section{Vacuum Stability}\label{sec:vacuum}

\begin{figure}
  \centering
    \includegraphics[width=0.9\linewidth]{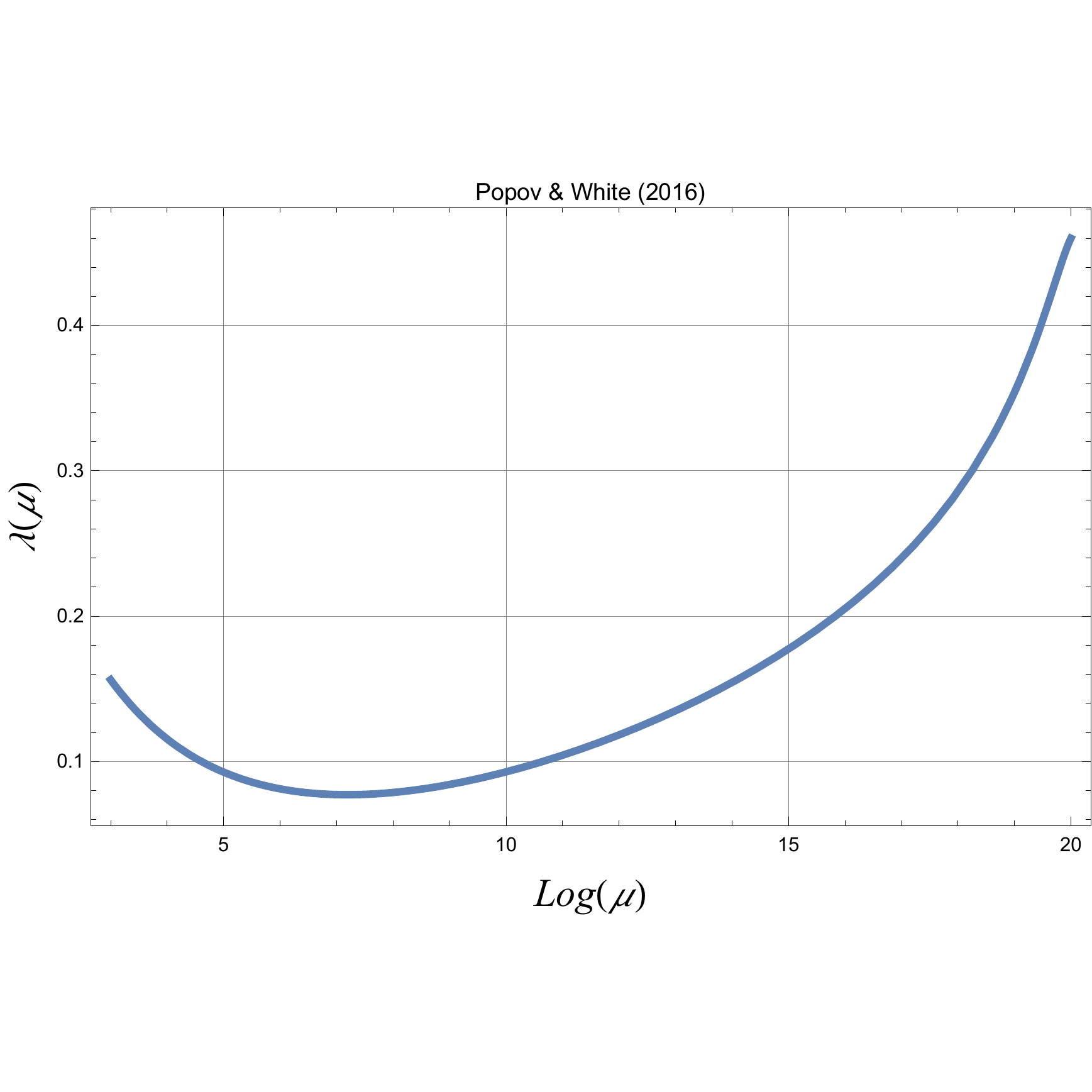} 
    \caption{\raggedright{}Vacuum stability achieved through improved running of the Higgs quartic coupling due to moderately large portal couplings ($g_{h\phi} =0.5$). This is compatible with low energy phenomology as well as explaining anomalies in the value of $(g-2)_\mu$, B decays and violation of lepton universality.}
    \label{Fig:stability}
\end{figure}
The recent discovery of a Higgs like Boson at $125$ GeV \cite{higgsdiscovery1,higgsdiscovery2} has led to the realization that the standard model vacuum is likely metastable \cite{metastable1,metastability2,metastability3,Xing:2011aa}. This is because the Higgs quartic coupling becomes negative at large energy scales. This has motivated research on how to improve the stability of the electroweak vacuum through addition to the standard model. There are two ways of getting a boost improve the stability of the vacuum 
\begin{itemize}
    \item improved running through corrections to the standard model RGEs. This has been done numerously in singlet extensions to the standard model \cite{singletrunning1,singletrunning2,singletrunning3,singletrunning4,singletrunning5}
    \item extended scalar sector, for adding a real singlet that acquires a vacuum expectation value. The result can be that the Higgs quartic coupling can be larger than its standard model value \cite{singletstability}.
\end{itemize}
The leptoquark cannot have a vacuum expectation value at zero temperature so the only option is through improved running \cite{Bandyopadhyay:2016oif}. The beta functions for the Higgs quartic coupling will get a correction compared to the standard model which we denote $\delta \lambda _1$ and $\delta \lambda _2$ for one and two loop contributions respectively. The one loop correction to the standard model RGE for the higgs quartic coupling receives contributions from the portal coupling only
\begin{equation}
    \frac{\beta _\lambda}{(4 \pi )^2} \ni \frac{6    }{16 \pi ^2} g_{H \phi }^2  \ .
\end{equation}
Note that the one loop correction is positive which makes a boost to stability possible in the first place. So if the portal coupling is large then the leptoquark contributes to improving the stability of the vacuum. This is true of both leptoquark representations that we consider. The two loop correction can be significant enough to justify a careful analysis. For the sake of simplicity the leptoquark coupling to one generation dominates. This is simply to show the structure of the two loop beta functions more clearly - our numerical calculations consider all generations. We can then write the dominant contribution as
\begin{align}
& \frac{\beta _\lambda }{(4 \pi )^4} \ni  \frac{1}{256 \pi ^4} \nonumber  \\ 
& \times \left\{\frac{6}{5} g_{H \phi }g_1^4-30 \lambda g_{H \phi }^2 + \frac{16}{5}g_1^2g_{H \phi }^2+64 g_3^2 g_{h \phi }^2 - 24 g_{H \phi }^3 \right. \nonumber  \\ 
& + g_{H \phi }^2 \lambda_{L,3}^2\left(24 + 9\lambda y_b^2-12y_b^4-12 y_\tau^4\right.\nonumber  \\ 
& \left. + 9 \lambda y_t^2-24y_\tau^2 y_t ^2-12 y_t^4 \right) \nonumber  \\
& \left. +g_{H \phi }^2 \lambda _{R,3}\left(12 - y_\tau ^4+9 \lambda y _t^2-24 y_t^2 y_\tau ^2-12 y_t^4 \right) \right\}
\end{align}
Typically this is strongly negative if the leptoquark couplings are large which drives up the value of $g_{h \phi } $ needed to make the higgs quartic coupling remain non-negative up to the unification scale. To achieve gauge coupling unification with all left handed leptoquark couplings set to zero, right handed leptoquark couplings need to be large. Also, one of the left handed leptoquark coupling constants has to be quite large to explain the violation of lepton universality. This means that several right handed leptoquark couplings have to be quite large to make gauge coupling unification manifest which pushes up the required value of $g_{h\phi }$ to a moderate value. In Fig. \ref{Fig:stability} we again the numerical package SARAH \cite{sarah} to run standard model parameters up to the scale of the leptoquark mass (which we choose to be $1$ TeV) and then run all parameters including the leptoquark parameters up to the a high scale. For a value of $g_\phi \sim 0.5$ the Higgs quartic coupling never goes negative. Finally we note that the extra fermions introduced to radiatively induce a neutrino mass at one loop only affect the running of the Higgs quartic at 2 loops.

\section{Leptoquarks and flavour anomalies}\label{sec:pheno}
In this section we consider the constraints on the parameter space due to both collider constraints and the proposal to have such a leptoquark explaining the afore mentioned anomalies. We should note that some doubt has been raised on whether this leptoquark representation can provide an explanation for the $R_K$ anomaly \cite{Becirevic:2016oho}. However, further inspection does indeed seem to confirm the original claim that this leptoquark can explain the apparent violation of lepton flavour universality \cite{Cai:2017wry} and recent work has indeed argued that it can indeed provide an explanation for the $R_{D^*}$ anomaly and the $(g-2)_\mu$ anomaly \cite{Dumont:2016xpj}. We will focus on briefly reviewing the parameter space proposed in \cite{explanation4} and find that there is tension between the parameter space they propose and unification. The parameter space discussed in \cite{Cai:2017wry} will has similar features to the parameter space in \cite{explanation4}. 

Let us begin by cataloguing the constraints from the anomalies. The constraints are written in terms of the leptoquark couplings written in the mass basis. To convert from the mass basis to the weak basis we make use of both the CKM matrix and the PMNS matrix as well as the relations
\begin{equation}
\lambda ^L _{ue} = U^T _u \lambda _L U_e , \ \lambda ^L _{d \nu} = U^T_d  \lambda _L, \ \lambda ^R _{ue} = V_u ^T \lambda _R V_e
\end{equation}
We will begin with the requirement that the leptoquark provide an explanation for anomalous B decays which gives the following condition on the leptoquark couplings
\begin{align}
    \lambda _{c \tau }  ^{L *}\lambda _{b \nu _ \tau }  ^{L} & = 3.5 \times 10 ^{-7} M^2 _\phi  \\
       \lambda _{c \tau }  ^{R *}\lambda _{b \nu _ \tau }  ^{L} & = -3 \times 10 ^{-8} M^2 _\phi 
\end{align}
These constraints cause two left handed couplings to be around $0.6$ and the right handed coupling $\lambda^R_{c \tau}$ about $0.05$. Since we require the trace of the right handed leptoquark coupling matrix to be significantly larger than the trace of the left handed leptoquark matrix this already constrains the parameter space somewhat. Next we turn to the anomalous measurement of $(g-2)_\mu$. This leads to the condition 
\begin{align}
&    (1+0.17 \ln 10^{-3}M_\phi) {\rm Re} (\lambda ^R _{c \mu } \lambda ^{L*} _{c \mu} ) \nonumber  \\ 
& + 20.7(1+1.06 \ln 10^{-3}M_\phi){\rm Re}(\lambda ^R _{t \mu } \lambda ^{L*} _{t \mu} ) \approx 8 \times 10^{-8} M_\phi ^2
\end{align}
This does not put much constraint on the right handed couplings as we can just tune the left handed couplings down.
The apparent violation of lepton universality, by contrast, puts the most severe constraint on the parameters as it inevitably leads to some coupling constants being large. Specifically the constraints are
\begin{align}
    & \sum _i |\lambda _{u _i \mu } ^L|^2 {\rm Re} \frac{\sum _j \lambda _{b\nu _j} ^L \lambda _{s \nu _j} ^{L *}}{ V_{tb} V_{ts}^*} - 1.74 |\lambda ^L _{t \mu} | ^2 \approx 1.25 \times 10^{-5} M_\phi ^2 \\
    & \frac{\sum _j \lambda _{b\nu _j} ^L \lambda _{s \nu _j} ^{L *}}{ V_{tb} V_{ts}^*} \approx  (1.87+0.45 i) \times 10^{-3} M_{\phi} 
\end{align}
 The second constraint is straightforward to satisfy because it requires that the sum of a set of left handed couplings must come to the small value of $0.076$ which is consistent with gauge coupling unification. Combining this with the first constraint though says that the square of two couplings, $|\lambda _{u  \mu}|^2+|\lambda _{c \mu }|^2$ must add to about $6.7$. This is obviously impossible to satisfy without making  at least one of these couplings very large. It can be more dangerous phenomenologically to have large leptoquark couplings to first generation quarks or leptons, so in reality this will lead to the coupling $|\lambda _{c \mu }|\approx 2.4$ for a leptoquark mass of around a TeV. This puts a very sharp constraint on the ability of this leptoquark to simultaneously give rise to gauge coupling unification and explain the apparent violation of lepton universality. Note that only the violation of lepton univerality requires a condition that conflicts with gauge coupling unification.

Let us next turn our attention to the phenomological constraints. We can get constraints on the leptoquark mass from reference \cite{pdg}. The lower bound for leptoquarks that decay into bottom quarks is $625$ GeV and the lower bound for leptoquarks decaying into muons with a significant branching ratio is $850$ GeV. The latter bound is more relevant if one is indeed desiring to explain an observed violation of lepton universality.
For convenience we will catalogue all other relevant collider constraints in a single list
\begin{align}
&-1.2 \times 10^{-6} M^2_\phi < {\rm Re}\frac{\sum _j \lambda _{b\nu _j} ^L \lambda _{s \nu _j} ^{L *} }{V_{tb} V_{ts}^*} < 2.25 \times 10^{-6} M^2_\phi  \nonumber \\
& \sqrt{|\lambda _{c \mu} ^L|^2|\lambda _{u \mu } ^R|^2+|\lambda ^L _{u \mu } |^2|\lambda ^R _{c \mu} |^2}< 1.2 \times 10^{-9} M_{\phi }^2  \nonumber \\
& |\lambda ^L_{c \mu } {\lambda ^L _{u \mu }}^*+\lambda ^R_{c \mu } {\lambda ^R _{u \mu }}^*| < 5.1 \times 10^{-8} \nonumber \\
& \sqrt{|\lambda ^L_{c \mu }|^2 +|\lambda ^L _{u \mu}|^2} <\frac{3.24 \times 10^{-3} M_\phi  }{\sqrt{1+0.39 \ln 10^{-3}M_\phi }} \nonumber \\
& |\lambda ^L _{t \mu } | <\frac{1.22 \times 10^{-3} M_\phi}{\sqrt{1+0.76 \ln 10^{-3}M_\phi }} \nonumber \\
 &  \Bigg[ \bigg| (1+0.17 \ln 10^{-3}M_\phi)  (\lambda ^R _{c \mu } \lambda ^{L*} _{c \mu} ) \nonumber \\ 
& + 20.7(1+1.06 \ln 10^{-3} M_\phi)(\lambda ^R _{t \mu } \lambda ^{L*} _{t \mu} ) - 0.015 \sum _i \lambda _{u_i \mu} ^{L*} \lambda _{u_i \tau } ^L \bigg|^2 \nonumber \\
&+(L \leftrightarrow R) \Bigg] ^{1/2} < 1.7 \times 10^{-8} M_\phi ^2
\end{align}
The constraints in the above list are, in order of their appearance
\begin{itemize}
    \item $B^- \rightarrow  K^- \nu \bar{\nu} $ and  $B^- \rightarrow  K^{-*} \nu \bar{\nu} $ decays. 
    \item Next two are from the bound on branching ratio of $D ^0 \rightarrow \mu ^+ \mu ^-$.  
    \item Next two are from constraints on the partial width of $Z \rightarrow \mu ^+ \mu ^-$.  
    \item Last is a bound on the branching ratio of $\tau \rightarrow \mu \gamma$ 
\end{itemize}

The constraint from the  $B^- \rightarrow  K^- \nu \bar{\nu} $ and  $B^- \rightarrow  K^{-*} \nu \bar{\nu} $ decays put no constraints on right handed couplings but can be satisfied with arbitrarily small left handed couplings.
The constraint from the branching ratio of $D ^0 \rightarrow \mu ^+ \mu ^-$ forces one $\Lambda ^R$ to be small.  It is satisfied if the left handed couplings are very small, the right handed couplings if they are equal can be $\sim 0.23$. Note that lepton universality means that the $\lambda ^L _{c\mu}$ coupling has to be $\approx 2.5$ which is fine if $\lambda ^L _{u \mu}$ is small. 
The constraints due to the partial width of $Z \rightarrow \mu ^+ \mu ^-$  leads to no constraint on the right handed couplings but is satisfied for arbitrarily small left handed couplings. Lepton universality requires that $\lambda ^L_{c \mu}$ left handed coupling to be large, $\sim 2.5$, but this is compatible with the constraint so long as other left handed couplings appearing in the constraint are suppressed which we desire anyway. 

The last constraint due to the branching ratio of $\tau \rightarrow \mu \gamma$  is the most dangerous if one desires the addition of a TeV scale leptoquark to lead to gauge coupling unification at two loops. There is a relative sign in the equation which means in principle that one can find a fine tuned region to the parameter space where a fortuitous cancellation occurs. But let us concentrate on the non-fine tuned region of parameter space. Recalling that the apparent violation of lepton universality leads to $\lambda^L _{c\mu}\sim 2.4$, this leads to $\lambda ^R_{\mu \nu_i}$ and $\lambda^R_{\tau \nu _i}$ needing to be have a value of approximately $\sim 0.08$. So we have 7 right handed couplings set to be small in principle. However, a loophole is just to set 3 of them to be very small. To give maximum freedom set right handed $\lambda ^R _{c \tau }$, $\lambda ^R _{c \mu }$, $\lambda ^R _{u \mu }$ to be small as well as $\lambda _{t \mu} ^R$. This will satisfy all of the above constraints without significant fine tuning.
Finally let us conclude this section by summarizing one of the recent works which called into question whether this leptoquark can explain the observed violation of lepton universality. Ref. \cite{Dumont:2016xpj} found that using this type of leptoquark to explain the $R_{D^(\star)}$ anomaly requires $\lambda _L ^{2j }$ 
\begin{align}
    \frac{\lambda_L^{3i } \lambda_R^{2 3}{}^\star}{2 M_{LQ}/GeV} &= -0.26 \times 2^{3/2} G_F V_{cb}  \\
  \frac{\lambda_L ^{33} \lambda_R^{23}{}^\star}{2 M_{LQ}^2/GeV^2} &= \pm 0.64 \times 2^{3/2} G_F V_{cb}
\end{align}
which is in serious tension with the requirement that $\lambda ^L_{c \mu}$ is large.

\section{discussion and conclusion}\label{sec:conclusion}
In this work we have considered some of the most intriguing experimental signatures that suggest a departure from standard model physics and attempted to see how a minimal explanation for them fits into the bigger picture of Unification. Improvement in vacuum stability via improved running is straightforward to realize. However, the attempt to explain a violation of lepton universality has some tension with using this leptoquark to achieve gauge coupling unification.

One way or another, we find that new physics is probably needed before the GUT scale. Achieving gauge coupling unification requires pushing leptoquark couplings up such that they are near the perturbativity bound at a high scale and neutrino masses cannot be achieved through this leptoquark alone. Both require additional paricle content (although if a tree level right handed neutrino mass is not forbidden by a discrete symmetry, the usual see-saw mechanism is of course sufficient). The single leptoquark model is insufficient to neutrino masses at any loop level. We proposed some minimal extensions by either introducing some gauge multiplets to generate a Majorana mass at two loops or including a heavy quark doublet with a hyper charge of $-5/6$ to generate such a mass at one loop. The latter is through d type mixing.
\acknowledgements
Graham White would like to acknowledge that this work was supported by the ARC Center of excellence for Particle Physics at the terra scale (CoePP). We would like to thank Ernest Ma and Peter Athron for helpful discussions and suggestions.


\begin{thebibliography}{50}

%\bibitem{example} A.~A.~Authorone and A.~A.~Authortwo, This is the Title, %Journ. Abbrev. {\bf 100}, 101 (2010)

\bibitem{explanation1}
%Freytsis, Marat, Zoltan Ligeti, and Joshua T. Ruderman. "Flavor models for B¯→ D (*) τ ν¯." Physical Review D 92.5 (2015): 054018.
M.~Freytsis, Z.~Ligeti and J.~T.~Ruderman, Flavor models for $B^- \rightarrow  D (*) \tau \nu ^¯$, Phys. Rev. D {\bf 92}, 054018 (2015).
\bibitem{explanation2}
% Murphy, Christopher W. "Vector leptoquarks and the 750 GeV diphoton resonance at the LHC." Physics Letters B 757 (2016): 192-198.
C.~W.~Murphy, Vector leptoquarks and the $750$ GeV diphoton resonance at the LHC, Phys. Lett. B {\bf 757}, 192 (2016).
\bibitem{explanation3}
%Fajfer, Svjetlana, and Nejc Košnik. "Vector leptoquark resolution of R K and RD (⁎) puzzles." Physics Letters B 755 (2016): 270-274.
S. Fajfer and N. Kosnik, Vector leptoquark resolution of RK and RD(*) puzzles, Phys. Lett. B {\bf 755}, 270 (2016).
\bibitem{explanation4}
%Bauer, Martin, and Matthias Neubert. "Minimal Leptoquark Explanation for the R D (*), R K, and (g− 2) μ Anomalies." Physical review letters 116.14 (2016): 141802.
M.~Bauer and M.~Neubert, Minimal Leptoquark Explanation for the RD(*), RK and $(g-2) _\mu$ anomalies, Phys. Rev. Lett. {\bf 116}, 141802 (2016).
\bibitem{explanation5}
%Doršner, I., et al. "Physics of leptoquarks in precision experiments and at particle colliders." arXiv preprint arXiv:1603.04993 (2016).
I.~Dorsner, et. al., Physics of leptoquarks in precision experiments and at particle colliders, [arXiv:1603.04993 [hep-ph]], (2016)
\bibitem{explanation6}
%Das, Diganta, et al. "Towards a unified explanation of R D (*), R K and (g− 2) μ anomalies in a left-right model with leptoquarks." Physical Review D 94.5 (2016): 055034.
D.~Das, et. al., Towards a unified eplanation of RD(*), RK and $(g-2)_\mu$ anomalies in a left-right model with leptoquarks, Phys. Rev. D, {\bf 94}, 055034 (2016)
\bibitem{explanation7}
%Li, Xin-Qiang, Ya-Dong Yang, and Xin Zhang. "Revisiting the one leptoquark solution to the $ R (D^{(\ ast)}) $ anomalies and its phenomenological implications." arXiv preprint arXiv:1605.09308 (2016).
X.~Li, Y.~Yang, and X.~Zhang, Revisiting the one leptoquark solution to the $ R (D^{(\ ast)}) $ anomalies and its phenomenological implications, [arXiv:1605.09308 [hep-ph]] (2016).
\bibitem{explanation8}
%Dumont, Béranger, Kenji Nishiwaki, and Ryoutaro Watanabe. "LHC constraints and prospects for S 1 scalar leptoquark explaining the B¯→ D (*) τ ν¯ anomaly." Physical Review D 94.3 (2016): 034001.
B.~Dumont, K.~Nishiwaki and R.~Watanabe, LHC constraints and prospects for $S_1$ scalar leptoquark explaining the $\bar B\ {\rm to} D^* \ tau \ \bar{\nu }$ anomaly, Phys. Rev. D, {\bf 94} 034001 (2016).
\bibitem{Calibbi:2015kma}
Lorenzo Calibbi, Andreas Crivellin, and Toshihiko Ota, Phys. Rev. Lett. {\bf 115}, 181801 (2015)
[hep-ph/1506.02661]
 \bibitem{Queiroz:2014pra} 
  F.~S.~Queiroz, K.~Sinha and A.~Strumia,
  %``Leptoquarks, Dark Matter, and Anomalous LHC Events,''
  Phys.\ Rev.\ D {\bf 91}, no. 3, 035006 (2015)
  doi:10.1103/PhysRevD.91.035006
  [arXiv:1409.6301 [hep-ph]].
  %%CITATION = doi:10.1103/PhysRevD.91.035006;%%
  %47 citations counted in INSPIRE as of 23 Dec 2016
  %\cite{Allanach:2015ria}
\bibitem{Allanach:2015ria} 
  B.~Allanach, A.~Alves, F.~S.~Queiroz, K.~Sinha and A.~Strumia,
  %``Interpreting the CMS $\ell^+\ell^- jj E\!\!\!\!/_{\rm T}$ Excess with a Leptoquark Model,''
  Phys.\ Rev.\ D {\bf 92}, no. 5, 055023 (2015)
  doi:10.1103/PhysRevD.92.055023
  [arXiv:1501.03494 [hep-ph]].
  \bibitem{Deppisch:2016qqd} 
  F.~F.~Deppisch, S.~Kulkarni, H.~Päs and E.~Schumacher,
  %``Leptoquark patterns unifying neutrino masses, flavor anomalies, and the diphoton excess,''
  Phys.\ Rev.\ D {\bf 94}, no. 1, 013003 (2016)
  doi:10.1103/PhysRevD.94.013003
  [arXiv:1603.07672 [hep-ph]].
  %%CITATION = doi:10.1103/PhysRevD.94.013003;%%
  %21 citations counted in INSPIRE as of 31 Dec 2016
\bibitem{anomalousBdecays1}
J. P. Lees et al. [BaBar Collaboration], Phys. Rev. Lett.
{\bf 109}, 101802 (2012) [arXiv:1205.5442 [hep-ex]].
\bibitem{anomalousBdecays2}
J. P. Lees et al. [BABAR Collaboration], Phys. Rev. D
{\bf 88}, 072012 (2013) [arXiv:1303.0571 [hep-ex]].
\bibitem{anomalousBdecays3} 
M. Huschle et al. [Belle Collaboration], arXiv:1507.03233
[hep-ex].
\bibitem{anomalousBdecays4}
R. Aaij et al. [LHCb Collaboration], arXiv:1506.08614
[hep-ex].
\bibitem{anomalousBdecays5}
B. Aubert et al. [BaBar Collaboration], Phys. Rev. Lett.
100, 021801 (2008) [arXiv:0709.1698 [hep-ex]].
\bibitem{anomalousBdecays6}
A. Bozek et al. [Belle Collaboration], Phys. Rev. D {\bf 82},
072005 (2010) [arXiv:1005.2302 [hep-ex]].
\bibitem{leptonuniversality}
R. Aaij et al. [LHCb Collaboration], Phys. Rev. Lett.
{\bf 113}, 151601 (2014) [arXiv:1406.6482 [hep-ex]]
\bibitem{muonmag}
M. Davier, A. Hoecker, B. Malaescu and Z. Zhang, Eur.
Phys. J. C {\bf 71}, 1515 (2011) [Eur. Phys. J. C 72, 1874
(2012)] [arXiv:1010.4180 [hep-ph]].
 \bibitem{Becirevic:2016oho} 
  D.~Bečirević, N.~Košnik, O.~Sumensari and R.~Zukanovich Funchal,
  %``Palatable Leptoquark Scenarios for Lepton Flavor Violation in Exclusive $b\to s\ell_1\ell_2$ modes,''
  JHEP {\bf 1611}, 035 (2016)
  doi:10.1007/JHEP11(2016)035
  [arXiv:1608.07583 [hep-ph]].
  %%CITATION = doi:10.1007/JHEP11(2016)035;%%
  %8 citations counted in INSPIRE as of 23 Dec 2016
 
  %%CITATION = doi:10.1103/PhysRevD.92.055023;%%
  %28 citations counted in INSPIRE as of 23 Dec 2016
%\cite{Cai:2017wry}
\bibitem{Cai:2017wry} 
  Y.~Cai, J.~Gargalionis, M.~A.~Schmidt and R.~R.~Volkas,
  %``Reconsidering the One Leptoquark solution: flavor anomalies and neutrino mass,''
  arXiv:1704.05849 [hep-ph].
  %%CITATION = ARXIV:1704.05849;%%
  %16 citations counted in INSPIRE as of 13 Jul 2017
\bibitem{E6ssm1}
  P.~Athron, S.~F.~King, D.~J.~Miller, S.~Moretti and R.~Nevzorov,
  %``The Constrained Exceptional Supersymmetric Standard Model,''
  Phys. Rev. D {\bf 80}, 035009 (2009)
  [arXiv:0904.2169 [hep-ph]].
  %%CITATION = doi:10.1103/PhysRevD.80.035009;%%
  %69 citations counted in INSPIRE as of 11 Oct 2016
\bibitem{E6ssm2}
P. Athron, et. al., Non-standard higgs decays in U (1) extensions of the MSSM, JHEP {\bf 2015} 1 (2015).
\bibitem{unification1}
H.~Murayama and T. Yanagida, Viable SU(5) GUT with light leptoquark bosons, Mod. Phys. Lett. A {\bf 7} 147 (1992)
\bibitem{unification2}
J.~ Hewett and T.~G.~Rizzo, Much ado about leptoquarks: A Comprehensive analysis, Phys. Rev. D {\bf 56} 5709 (1997)
\bibitem{unification3}
%%NOTE OLEG: This is an example where neutrino masses are included in a unification paper
I.~Dorsner and P.~Fileviez Pérez, Unification without supersymmetry: Neutrino mass, proton decay and light leptoquarks, Nucl. Phys. B {\bf 723} 53 (2005).
\bibitem{unification4}
  J.~Hewett and T.~G.~Rizzo, Don’t stop thinking about leptoquarks: Constructing new models, Phys. Rev. D {\bf 58} 055005 (1998).
\bibitem{LQmass2}
K.~S.~Babu and J. Julio, Two-loop neutrino mass generation through leptoquarks, Nucl. Phys. B {\bf 841} 130 (2010).
\bibitem{LQmass3}
L.~Jin, R.~Tang, and F.~Zhang, A three-loop radiative neutrino mass model with dark matter, Phys. Lett. B {\bf 741} 163 (2015).
\bibitem{LQmass4}
A.~Ahriche, K.~L.~McDonald S.~Nasri, Three-Loop Neutrino Mass Models at Colliders. [arXiv:1505.04320 [hep-ph]] (2015).
\bibitem{LQmass5}
F.~F.~Deppisch, et al., Leptoquark patterns unifying neutrino masses, flavor anomalies and the diphoton excess, [arXiv:1603.07672 [hep-ph]] (2016).
\bibitem{LQmass6}
P.~F.~Pérez, et al., Leptoquarks and neutrino masses at the LHC, Nucl. Phys. B {\bf 819}, 139 (2009).
\bibitem{LQmass7}
U.~Mahanta, Neutrino masses and mixing angles from leptoquark interactions, Phys. Rev. D {\bf 62}, 073009 (2000).
\bibitem{LQmass8}
C.~Chua, X.~He, and W.~Y.~P.~Hwang, Neutrino mass induced radiatively by supersymmetric leptoquarks, Phys. Lett. B {\bf 479} 224 (2000).

%\cite{Dorsner:2017wwn}
\bibitem{LQmass9} 
  I.~Doršner, S.~Fajfer and N.~Košnik,
  %``Leptoquark mechanism of neutrino masses within the grand unification framework,''
  arXiv:1701.08322 [hep-ph].
  %%CITATION = ARXIV:1701.08322;%%
  %\cite{Deppisch:2016qqd}
\bibitem{LQmass10} 
  F.~F.~Deppisch, S.~Kulkarni, H.~Päs and E.~Schumacher,
  %``Leptoquark patterns unifying neutrino masses, flavor anomalies, and the diphoton excess,''
  Phys.\ Rev.\ D {\bf 94}, no. 1, 013003 (2016)
  doi:10.1103/PhysRevD.94.013003
  [arXiv:1603.07672 [hep-ph]].
  %%CITATION = doi:10.1103/PhysRevD.94.013003;%%
  %23 citations counted in INSPIRE as of 01 Apr 2017
  %\cite{Babu:2010vp}
\bibitem{LQmass11} 
  K.~S.~Babu and J.~Julio,
  %``Two-Loop Neutrino Mass Generation through Leptoquarks,''
  Nucl.\ Phys.\ B {\bf 841}, 130 (2010)
  doi:10.1016/j.nuclphysb.2010.07.022
  [arXiv:1006.1092 [hep-ph]].
  %%CITATION = doi:10.1016/j.nuclphysb.2010.07.022;%%
  %52 citations counted in INSPIRE as of 01 Apr 2017
  %\cite{FileviezPerez:2008dw}
\bibitem{LQmass12} 
  P.~Fileviez Perez, T.~Han, T.~Li and M.~J.~Ramsey-Musolf,
  %``Leptoquarks and Neutrino Masses at the LHC,''
  Nucl.\ Phys.\ B {\bf 819}, 139 (2009)
  doi:10.1016/j.nuclphysb.2009.04.009
  [arXiv:0810.4138 [hep-ph]].
  %%CITATION = doi:10.1016/j.nuclphysb.2009.04.009;%%
  %26 citations counted in INSPIRE as of 01 Apr 2017
\bibitem{Cox:2016epl} 
  P.~Cox, A.~Kusenko, O.~Sumensari and T.~T.~Yanagida,
  %``SU(5) Unification with TeV-scale Leptoquarks,''
  arXiv:1612.03923 [hep-ph].
  %%CITATION = ARXIV:1612.03923;%%

\bibitem{sarah}
F.~Staub, SARAH 4: A tool for (not only SUSY) model builders, Comp. Phys. Comm. {\bf 185} 1773 (2014).
\bibitem{higgsdiscovery1}
S. Chatrchyan {\it et al.} [CMS Collaboration], Phys. Lett. B {\bf 716}, 30 (2012).
\bibitem{higgsdiscovery2}
G. Aad {\it et al.} [ATLAS Collaboration], Phys. Lett. B {\bf 716}, 1 (2012).
\bibitem{metastable1}
S. Alekhina,  A. Djouadib, and  S. Moch, Phys. Lett. B {\bf 716}, 214 (2012).

\bibitem{metastability2}
J. Elias-Miro, J. R. Espinosa, G. F. Giudice, G. Isidori, A. Riotto and
A. Strumia, [hep-ph/1112.3022]
  
\bibitem{metastability3}
M. Holthausen, K. S. Lim and M. Lindner, [hep-ph/1112.2415].
  %\cite{Xing:2011aa}
\bibitem{Xing:2011aa} 
  Z.~z.~Xing, H.~Zhang and S.~Zhou,
  %``Impacts of the Higgs mass on vacuum stability, running fermion masses and two-body Higgs decays,''
  Phys.\ Rev.\ D {\bf 86}, 013013 (2012)
  doi:10.1103/PhysRevD.86.013013
  [arXiv:1112.3112 [hep-ph]].
  %%CITATION = doi:10.1103/PhysRevD.86.013013;%%
  %96 citations counted in INSPIRE as of 23 Dec 2016

\bibitem{singletrunning1}

M.  Gonderinger,  Y.  Li,  H.  Patel  and  M.  J.  Ramsey-Musolf,  JHEP  {\bf 2010}, 1001  (2010).   
\bibitem{singletrunning2}
O.  Lebedev  and  H.  M.  Lee,  Eur.  Phys.  J.  C  {\bf 71} 1821  (2011) 
\bibitem{singletrunning3}
M. Kadastik, K. Kannike, A. Racioppi and M. Raidal, [hep-ph/1112.3647];
\bibitem{singletrunning4}
M. Gonderinger, H. Lim and M. J. Ramsey-Musolf, [hep-ph/1202.1316]; 
\bibitem{singletrunning5}
C. -S. Chen and Y. Tang [hep-ph/1202.5717]
\bibitem{singletstability}
J. Elias-Miró, J.~R. Espinosa, G.~F. Giudice,  {\it et al.} JHEP {\bf 2012}, 31 (2012).
 \bibitem{Bandyopadhyay:2016oif} 
  P.~Bandyopadhyay and R.~Mandal,
  %``Vacuum stability in extended standard model with leptoquark,''
  arXiv:1609.03561 [hep-ph].
  %%CITATION = ARXIV:1609.03561;%%
  %\cite{Aitchison:2005cf}
\bibitem{pdg}
C. Patrignani et al. (Particle Data Group), Chin. Phys. C, 40, 100001 (2016) and 2017 update.
\bibitem{weinberg}
S. Weinberg, Phys. Rev. Lett. 43, 1566 (1979).

\bibitem{Dumont:2016xpj} 
  B.~Dumont, K.~Nishiwaki and R.~Watanabe,
  %``LHC constraints and prospects for $S_1$ scalar leptoquark explaining the $\bar B \to D^{(*)} \tau \bar\nu$ anomaly,''
  Phys.\ Rev.\ D {\bf 94}, no. 3, 034001 (2016)
  doi:10.1103/PhysRevD.94.034001
  [arXiv:1603.05248 [hep-ph]].
  %%CITATION = doi:10.1103/PhysRevD.94.034001;%%
  %15 citations counted in INSPIRE as of 23 Dec 2016
 

\bibitem{Aitchison:2005cf} 
  I.~J.~R.~Aitchison,
  %``Supersymmetry and the MSSM: An Elementary introduction,''
  hep-ph/0505105.
  %%CITATION = HEP-PH/0505105;%%
  %157 citations counted in INSPIRE as of 24 Dec 2016
  %\cite{Cox:2016epl}
\end{thebibliography}
\end{document}